# Distributed MAC Protocol Supporting Physical-Layer Network Coding

Shiqiang Wang, Student *Member, IEEE*, Qingyang Song, *Member, IEEE*,
Xingwei Wang, and Abbas Jamalipour, *Fellow, IEEE*

**Abstract**—Physical-layer network coding (PNC) is a promising approach for wireless networks. It allows nodes to transmit simultaneously. Due to the difficulties of scheduling simultaneous transmissions, existing works on PNC are based on simplified medium access control (MAC) protocols, which are not applicable to general multi-hop wireless networks, to the best of our knowledge. In this paper, we propose a distributed MAC protocol that supports PNC in multi-hop wireless networks. The proposed MAC protocol is based on the carrier sense multiple access (CSMA) strategy and can be regarded as an extension to the IEEE 802.11 MAC protocol. In the proposed protocol, each node collects information on the queue status of its neighboring nodes. When a node finds that there is an opportunity for some of its neighbors to perform PNC, it notifies its corresponding neighboring nodes and initiates the process of packet exchange using PNC, with the node itself as a relay. During the packet exchange process, the relay also works as a coordinator which coordinates the transmission of source nodes. Meanwhile, the proposed protocol is compatible with conventional network coding and conventional transmission schemes. Simulation results show that the proposed protocol is advantageous in various scenarios of wireless applications.

✦

## 1 INTRODUCTION

RECENT advances in network coding brings performance improvement for wireless relaying networks [1]. The basic idea of network coding is to encode packets at a relay before forwarding. By this means, the bandwidth allocated to each node can be utilized more efficiently. Compared with conventional relaying/routing schemes, the network throughput, end-to-end delay and network reliability can be improved with network coding. In conventional network coding (CNC) schemes [2], [3], [4], the relay encodes packets after receiving them in separate communication phases. Physical-layer network coding (PNC) [5], [6], [7], [8], [9] encodes packets through simultaneous transmissions. According to electromagnetic theories, simultaneously transmitted electromagnetic waves superpose in space, and the relay obtains an encoded version of the original packets from this superposed signal. PNC further reduces the number of required communication phases, and hence increases throughput. The underlying idea of PNC is that wireless interference should not always be considered harmful. Fig. 1 illustrates the conventional relaying, CNC and PNC schemes.

It is shown in Fig. 1 that for the Alice-and-Bob topology, PNC requires half of the time compared with conventional relaying and 2/3 of the time compared with CNC, to transmit one packet from each of the nodes *A* and *B*. This reduction of transmission time corresponds to a throughput gain of 2 over conventional relaying and 1.5 over CNC. The encoding function $C_{\text{CNC}}(\cdot)$ of CNC is usually an XOR operation for the Alice-and-Bob topology. The encoding function $C_{\text{PNC}}(\cdot)$ of PNC can be either a multiplicative factor that amplifies the signal or an operator that maps the superposed signal to a series of bits representing the encoded packet [8]. The destination can decode the packet it intends to receive from the encoded packet (or the amplified superposed signal) forwarded by the relay *R*, because it is aware of the packet sent by itself.

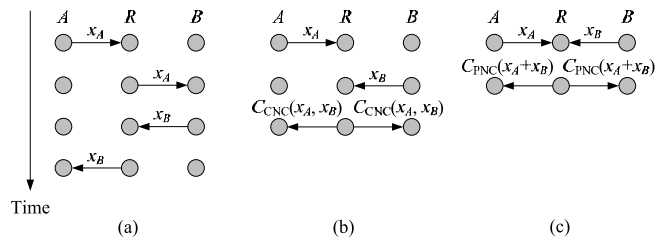

Fig. 1. Three relaying schemes: (a) conventional relaying, (b) CNC, (c) PNC. Nodes *A* and *B* exchange packets through the relay *R*. $C_{\text{CNC}}(\cdot)$ and $C_{\text{PNC}}(\cdot)$ respectively represent the encoding function of CNC and PNC.

• S. Wang is with the School of Information Science and Engineering, Northeastern University, Shenyang 110819, P. R. China, and also with the Dept. of Electrical and Electronic Engineering, Imperial College London, SW7 2AZ, United Kingdom. E-mail: shiqiang.wang11@imperial.ac.uk.
• Q. Song (Corresponding Author) and X. Wang are with the School of Information Science and Engineering, Northeastern University, Shenyang 110819, P. R. China. E-mail: songqingyang@ise.neu.edu.cn, wangxw@mail.neu.edu.cn.
• A. Jamalipour is with the School of Electrical and Information Engineering, University of Sydney, NSW, 2006, Australia. E-mail: a.jamalipour@ieee.org.

One major issue that arises in wireless networks is how wireless terminals access the channel. Medium access control (MAC) protocols have been developed to coordinate channel access. The time division multiple access (TDMA) or frequency division multiple access (FDMA) schemes require central scheduling, which is generally difficult to implement in distributed wireless networks. Hence, random access mechanisms, such as carrier sense multiple access (CSMA), have been widely adopted in wireless local area networks (WLANs) and wireless ad hoc networks [10], [11]. However, compared with centrally scheduled MAC protocols, random access MAC protocols have higher complexity. Therefore, most related works on PNC in the literature assume a TDMA-like MAC layer [5], [6], [7], [8], [12], [13], [14], [15].

The focus of this paper is to develop a practical distributed MAC protocol that supports PNC and is based on random access strategies. We refer to the proposed MAC protocol as PNC-MAC in our further discussions. PNC-MAC extends the CSMA-based IEEE 802.11 MAC protocol [11] and works in multi-hop wireless networks with arbitrary topologies. Nodes randomly access the channel as in the conventional IEEE 802.11 MAC. When there is an opportunity to perform PNC, PNC-MAC coordinates the source nodes to transmit simultaneously. The basic idea of PNC-MAC is to encourage instructive interference which can be used for PNC and, at the same time, avoid destructive interference that may result in packet losses. In the cases where PNC is not applicable, PNC-MAC automatically switches back to the CNC or conventional relaying schemes. When using CNC, PNC-MAC employs reliable broadcasting as proposed in [16].

The remainder of this paper is organized as follows. Section 2 summarizes the related work. Section 3 discusses some detailed issues on the problem we address in this paper. Section 4 introduces the basic principles of the proposed PNC-MAC protocol. Section 5 describes the queuing method of packets and how to select the appropriate relaying method (i.e. PNC, CNC or conventional relaying). Details on framing are discussed in Section 6. Section 7 evaluates the performance of PNC-MAC through simulations and Section 8 draws conclusions.

## 2 RELATED WORK

The distributed control of communication networks remains a challenging area [1], [4], [8], [17], [18], [20], [21], [22], [23], [24], [25], [26], [27]. Recently, efforts on developing distributed MAC protocols that support PNC have been made, to make PNC applicable in practical systems. A basic MAC protocol for simple topologies using PNC was proposed in [17], which was implemented on a software-defined radio prototype. In that protocol, a tail, which contains the same information as the header, is added to the packet, so that the header information can be successfully decoded from the non-superposed part of the superposed packet. However, it did not consider random access issues, and hence, it is difficult to be applied to more sophisticated topologies. In [18], a cooperative protocol for PNC was proposed, which allows partial packet superposition to make PNC feasible for distributed wireless networks with one relay and two sender-destination pairs. A theoretical MAC protocol for PNC using the abstract MAC layer specification is proposed in [19]. In [20], a distributed MAC protocol that supports rate-adaptive cooperative transmissions (including PNC) was proposed, which optimizes the data rates within a single hop. Our work differs from [20] in the sense that we consider queuing issues and the interaction between nodes that are connected in a multi-hop fashion. These issues are significant for multi-hop networks, because not every node has always a packet to send.

Self interference cancellation (SIC) schemes, which also make use of simultaneous transmissions, were incorporated with distributed MAC protocols in [21] and [22], which work in general wireless networks. The difference between PNC and SIC is that, with PNC, the superposed signal is mapped to a signal representing a coded packet; and with SIC, a packet is extracted from the superposed signal based on the node's knowledge of its previously stored packets.

Distributed MAC protocols which enhance the performance of CNC were also investigated in [16], [23], and [24].

## 3 PROBLEM STATEMENT

In this section, we detail the problem that is addressed in this paper.

### 3.1 Type and Number of Packets to be Encoded

In this paper, when using PNC, the simultaneously transmitting nodes transmit original source packets (and not a coded version of the packets), and we restrict the number of packets that are encoded to two. This restriction simplifies the protocol design and relaxes the hardware requirements, because sophisticated optimization, synchronization, or interference cancellation strategies may be required when the encoding packet number exceeds two [15]. When using CNC, we do not impose restrictions on the number of packets that are encoded.

### 3.2 Type of Flows

In Fig. 1, node $A$ intends to send packets to node $B$, and conversely, node $B$ intends to send packets to node $A$. Hence, there exist data *flows* $A \rightarrow B$ and $B \rightarrow A$. If the flows $A \rightarrow B$ and $B \rightarrow A$ both exist, we can also say that there exists a *bidirectional flow* $A \leftrightarrow B$. It has been shown in [6] and [7] that PNC is beneficial for bidirectional flows. For unidirectional flows, the destination node generally has to overhear the packet sent by nearby source node through opportunistic listening, so that the coded packet can be decoded at the destination. However, the effective overhearing range of PNC is low compared with CNC schemes, because simultaneous transmissions introduce additional interference [15]. Another method of using PNC with unidirectional flows was proposed in [6], but that approach is similar with the SIC technique, and a MAC protocol supporting SIC was proposed in [22]. For these reasons, we focus on bidirectional flows in this paper. Application examples of bidirectional flows include peer-to-peer file exchange, video communications etc. Also, for a network with multiple unidirectional flows, bidirectional communication may be formed up where unidirectional flows in different directions overlap.

### 3.3 Signal Analysis and Method of Performing PNC

According to the aforementioned discussions, the Alice-and-Bob topology shown in Fig. 1 is a typical example of our problem under consideration, in which two packets are encoded and a bidirectional flow is considered.

Let $h_{ij}$ denote the channel gain from node $i$ to node $j$. In the first communication phase, the signal received at the relay is

$$y_R = h_{AR}x_A + h_{BR}x_B + z_n, \quad (1)$$

where $x_A$ and $x_B$ are the signals transmitted by nodes $A$ and $B$, respectively; and $z_n$ is noise.

In the second communication phase, the signals received at nodes $A$ and $B$ are respectively

$$y_A = h_{RA}C_{PNC}(y_R) + z_n, \quad (2)$$
$$y_B = h_{RB}C_{PNC}(y_R) + z_n. \quad (3)$$

The encoding function $C_{PNC}(\cdot)$ can be a linear function when using the amplify-and-forward (AF) method of PNC or a non-linear function when using the denoise-and-forward (DNF) method or the two-phase decode-and-forward (DF) method of PNC [8]. Different encoding functions have different synchronization requirements and decoding operations, which also result in different performance. The proposed PNC-MAC protocol can work with any encoding function that is suitable for PNC. In our simulations in Section 7, we use the DNF method.

### 3.4 Challenges for PNC-MAC

The major challenges of developing PNC-MAC are outlined as follows.

1. Coordinating simultaneous transmissions: When performing PNC, the source nodes carrying the packets to be encoded have to transmit simultaneously. Hence, a coordination scheme needs to be developed to coordinate the simultaneous transmissions that are used for PNC and, at the same time, avoid the destructive collisions.
2. Checking for PNC opportunity: Each node has to be aware of whether it has the opportunity to perform PNC before transmitting a packet, because if the node has this opportunity, it has to transmit simultaneously with the other node that transmits the packet to be coded together.
3. Compatibility with other relaying schemes: Because performing PNC may not always be possible and advantageous, the proposed protocol should be compatible with CNC and conventional relaying schemes.

The aforementioned challenges are the major design considerations of the proposed PNC-MAC protocol and are resolved in this paper.

## 4 BASIC PRINCIPLES OF PNC-MAC

In this section, we discuss the basic principles of the proposed PNC-MAC protocol. PNC-MAC modifies the request-to-send/clear-to-send (RTS/CTS) based IEEE 802.11 MAC protocol, to support PNC. In PNC-MAC, each node notifies its queue status to its neighboring nodes by adding a few bytes of control information to the data and acknowledgement (ACK) frames (which will be described in detail in Section 6). In this paper, we also assume a proactive routing protocol, in which nodes are aware of the network topology within at least two-hop range. When a node (denoted by $R$) senses according to the stored queue status (the queuing details will be discussed in Section 5) and routing information that there is an opportunity for two of its neighboring nodes (denoted by $A$ and $B$, respectively) to exchange packets through PNC, with $R$ as the relay, node $R$ performs channel access with the CSMA strategy and coordinates the nodes $A$ and $B$ to send packets. The detailed timing diagram is shown in Fig. 2. In Fig. 2 and the following discussions, $T_{SIFS}$ and $T_{DIFS}$ respectively denote the short inter-frame space (SIFS) and the distributed inter-frame space (DIFS) as defined in the IEEE 802.11 standard [11]; $T_{PHY-Hd}$, $T_{MAC-Hd}$, $T_{RTS-PNC}$, $T_{CTS}$, $T_{CO-PNC}$, $T_{ACK}$ and $T_{ACK-PNC}$ respectively denote the time length of the physical-layer header, MAC header, RTS-PNC, CTS, CO-PNC, ACK and ACK-PNC frame; $T_{DATA}(N)$ denotes the time length of the data frame sent by node $N$.

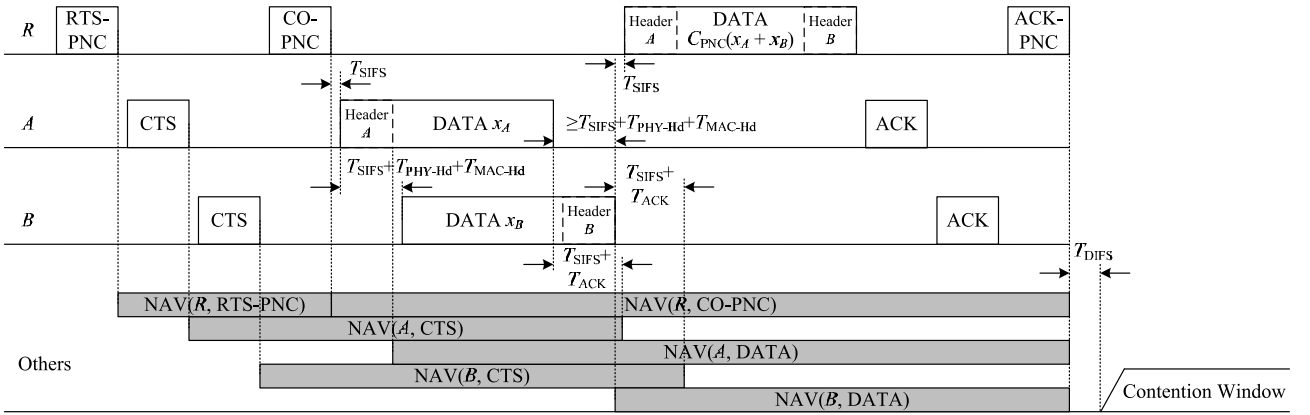

Fig. 2. Timing diagram of packet exchange using PNC.

As shown in Fig. 2, unlike conventional transmissions, a round of packet exchange using PNC is initiated by the relay. During the packet exchange process, the relay also acts as a coordinator that coordinates packet transmission. We focus on the timing of PNC-MAC in this section. Issues on how the queue statuses are transmitted and stored, as well as how to judge whether a node should initiate PNC or conventional transmission, will be discussed in Section 5.

### 4.1 Packet Exchange Process using PNC

When node $R$ senses that there is opportunity to perform PNC, it sends an RTS-PNC frame, which contains the addresses of the two source/destination nodes $A$ and $B$ and the address of node $R$. The node that has a shorter packet to send (which can be known from the queue status stored at $R$) is set as node $A$, for reasons described in the next paragraph. After receiving the RTS-PNC frame, the nodes $A$ and $B$ separately respond to $R$ with CTS frames. When node $R$ successfully receives both CTS frames, it sends a coordination (CO-PNC) frame, to coordinate packet transmissions of the nodes $A$ and $B$.

After receiving CO-PNC, node $A$ starts data transmission after time $T_{SIFS}$, node $B$ starts data transmission after time $2T_{SIFS} + T_{PHY-Hd} + T_{MAC-Hd}$. This process requires that nodes $A$ and $B$ are synchronized. The development of an explicit synchronization scheme is beyond the scope of this paper. However, we assume that the relay can estimate the timing difference between the two source nodes from the CTS frames it has received. With this information, the relay can send a compensation time to a specific source node (e.g. $A$) in the CO-PNC frame, so that node $A$ can adjust its timer to synchronize with node $B$ and ensure that PNC can be successfully performed. The accuracy requirement of the synchronization depends on the method of performing PNC. High synchronization accuracy is required when using the synchronous DNF or two-phase DF schemes, and only coarse synchronization is needed when using the AF or asynchronous DNF [9] methods.

Meanwhile, the data frame of node $B$ is in a bit-reversed order, i.e. the tail of the data frame is transmitted at first and the header at last. Because the packet sent by node $A$ is not longer than the packet sent by node $B$, there exists at least a time of $T_{SIFS} + T_{PHY-Hd} + T_{MAC-Hd}$ during which the data frame from node $B$ is not interfered. The time difference between the two data frames ensures that the relay $R$ can successfully decode the headers of both data frames. Decoding the headers is significant for judging whether the intended packets are superposed and for updating node $R$ with the most recent queue statuses of nodes $A$ and $B$. The strategy of adding a time difference between data frames is similar with the strategy proposed in [17].

After the relay $R$ receives the simultaneously transmitted and partly superposed signal, it performs the coding operation $C_{PNC}(\cdot)$ to the superposed part of the signal. The resulting coded packet is forwarded to the nodes $A$ and $B$. When node $A$ or $B$ receives the coded packet, it attempts to extract the packet, which it intends to receive, from the coded packet. If successful, an ACK frame is transmitted in the order as shown in Fig. 2. After the relay $R$ receives the ACK frames, an ACK-PNC frame is generated which contains the address(es) of the source node(s) to be acknowledged. After receiving ACK-PNC, each acknowledged source node flushes the packet, which it has just sent, from the queue.

According to the aforementioned discussions, when performing PNC, the packet that is encoded with PNC is directly forwarded after reception and at most one superposed signal is stored in the relay's buffer. This design is because 1) storing the superposed signal requires a relatively large buffer space; 2) when using the AF or DNF schemes of PNC, the relay cannot judge whether the superposed packet has been received successfully, and hence, only the destination can send acknowledgement to the corresponding source nodes. It follows that the proposed PNC-MAC protocol is suitable for various schemes of PNC.

### 4.2 Handling Exceptions

We discussed in Section 4.1 the basic process of exchanging packets using PNC. In this subsection, we discuss the exceptional cases caused by frame loss, and introduce how PNC-MAC handles these cases.

#### 4.2.1 RTS-PNC Received by A or B but No Packet to Send

In normal operation, node $R$ may only request nodes $A$ and $B$ to exchange packets using PNC when it infers that they have packets to exchange, according to the queue statuses of nodes $A$ and $B$ that are stored at node $R$. However, the queue statuses at node $R$ may not be up-to-date, due to the possible loss of frames carrying queue status information. To consider this case, we indicate in the CTS frame whether the source node has packets to send upon receiving RTS-PNC. When a node (e.g. $A$) receives an RTS-PNC from its neighboring node $R$ but finds that it has no packet, whose next hop is $R$ and second hop is $B$, to send, it still responds to node $R$ with a CTS but indicates in the frame that it has no packet to send.

#### 4.2.2 No CTS Received by R or All CTSs Received by R Indicate No Packet to Send

When the relay $R$ has received no CTS or every CTS it has received indicates that the source node has no packet to send, the relay $R$ does not send CO-PNC and a new round of channel contention is initiated.

### 4.2.3 Only One CTS Received by R Indicates that the Node Has Packet to Send

This subsection considers the following situations: 1) only one CTS, which indicates that the node has packet to send, is received; 2) two CTSs are received but one of them indicates that the node has no packet to send. In both situations, node $R$ notifies (with CO-PNC) the node, which has packet to send and successfully responded to $R$ with CTS, to transmit its packet. However, because only one node is transmitting, PNC cannot be performed and the packet is acknowledged by node $R$ directly after correct reception, as illustrated in Fig. 3.

As shown in Fig. 3, the start time of data frame transmission is the same as the ordinary case described in Section 4.1, i.e. the data frame from node $B$ is transmitted with an additional delay. Although the additional delay is unnecessary in this case, this design simplifies the timing and network allocation vector (NAV) setting, while only introducing a small loss of bandwidth efficiency particularly when the packet size is large.

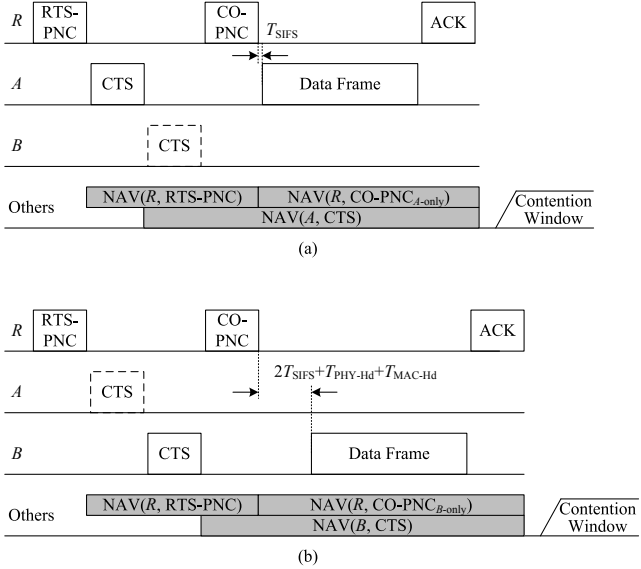

Fig. 3. The case where only one node transmits: (a) node $A$ transmitting, (b) node $B$ transmitting.

### 4.2.4 Erroneous Superposed Data Frame Received by R

When using the two-phase DF scheme of PNC, the superposed data frame received by the relay can be checked for errors [8]. When using the other schemes, we can check whether the header contains error if a frame check sequence (FCS) is added to the header. We can also predict whether there exists interference that can lead to an error by measuring the received signal strength [17] or by correlating the received signal with the known preamble [21]. If using any method, an error in the superposed data frame is detected, the relay $R$ does not forward the coded packet and initiates a new channel contention round. The nodes $A$ and $B$ wait for the expiration of the NAV timer, as shown in Fig. 2, and also contend the channel afterwards. Waiting for NAV expiration is necessary for nodes $A$ and $B$, because if node $A$, for instance, did not receive the packet sent by node $R$, there could be two reasons: 1) node $R$ did not send the packet, or 2) the packet from node $R$ was interfered. In the latter case, node $B$ may still be able to receive the packet from node $R$. Nodes $A$ and $B$ cannot distinguish which case has happened, until the expiration of the NAV, at which time the ACK signaling (if any) has been completed.

### 4.2.5 Erroneous Data Packet Decoded by A or B

When the data packet received and decoded by the destination node is erroneous, the destination does not send ACK and the corresponding source node is not acknowledged. In this case, the packet remains in the queue of the source node and will be retransmitted or dropped depending on the retry counter, as in the IEEE 802.11 standard [11]. If the relay $R$ has received no ACK from either of the nodes $A$ and $B$, it does not send ACK-PNC.

## 4.3 NAV Setting

When nodes other than $A$, $B$ and $R$ receive the frames sent during the packet exchange process, they set their NAV timers and remain silent (i.e. do not transmit) for the time specified in the NAVs. This strategy is called virtual carrier-sensing which avoids collisions in the presence of hidden terminals [11]. When exchanging packets using PNC in the proposed PNC-MAC protocol, the nodes $A$, $B$ and $R$ respectively set the NAV in two stages, to avoid unnecessary channel occupation. The length of the NAV is carried in the duration field of the frame. Different frames carry different NAV length, as shown in Figs. 2 and 3. Detailed description of the NAV lengths in different types of frames is presented in below.

### 4.3.1 RTS-PNC

The NAV length in RTS-PNC is set so that the relay $R$ occupies the channel until CO-PNC has been sent and a new NAV is set. More specifically, the NAV length is

$$T_{\text{NAV}}(R, \text{RTS-PNC}) = 3T_{\text{SIFS}} + 2T_{\text{CTS}} + T_{\text{CO-PNC}}. \quad (4)$$

### 4.3.2 CTS

When a node has no packet to send, it sets the NAV length in the CTS frame to $T_{\text{NAV}}(A, \text{CTS}_{\text{no-pk}}) = T_{\text{NAV}}(B, \text{CTS}_{\text{no-pk}}) = 0$. Otherwise, node $A$ (or $B$) sets the NAV length to cover the transmission time supposing only node $A$ (or $B$) transmits, as discussed in Section 4.2.3. Specifically, node $A$ sets the NAV length to

$$T_{\text{NAV}}(A, \text{CTS}) = 4T_{\text{SIFS}} + T_{\text{CTS}} + T_{\text{CO-PNC}} + T_{\text{DATA}}(A) + T_{\text{ACK}}, \quad (5)$$

and node $B$ sets the NAV length to

$$T_{\text{NAV}}(B, \text{CTS}) = 4T_{\text{SIFS}} + T_{\text{CO-PNC}} + T_{\text{PHY-Hd}} + T_{\text{MAC-Hd}} + T_{\text{DATA}}(B) + T_{\text{ACK}}. \quad (6)$$

When nodes $A$ and $B$ both transmit, new NAVs will be set in the headers of the data frames sent by $A$ and $B$ to cover the total time used for packet exchange with PNC, which will be discussed in Section 4.3.4.

### 4.3.3 CO-PNC

The NAV length in CO-PNC depends on which node(s) will transmit data, which can be known from the received CTS frames, as discussed in Section 4.2, and covers the remaining time used for data exchange. If only node $A$ transmits data, the relay $R$ sets the NAV length in CO-PNC to

$$T_{\text{NAV}}(R, \text{CO-PNC}_{A\text{-only}})$$
$$= T_{\text{NAV}}(A, \text{CTS}) - 2T_{\text{SIFS}} - T_{\text{CTS}} - T_{\text{CO-PNC}}. \quad (7)$$

If only node $B$ transmits data, the NAV length is set to

$$T_{\text{NAV}}(R, \text{CO-PNC}_{B\text{-only}}) = T_{\text{NAV}}(B, \text{CTS}) - T_{\text{SIFS}} - T_{\text{CO-PNC}}. \quad (8)$$

If both nodes $A$ and $B$ transmit data, the NAV length is set to

$$T_{\text{NAV}}(R, \text{CO-PNC})$$
$$= 2(T_{\text{NAV}}(B, \text{CTS}) - 2T_{\text{SIFS}} - T_{\text{CO-PNC}} - T_{\text{ACK}})$$
$$+ 3T_{\text{SIFS}} + 2T_{\text{ACK}} + T_{\text{ACK-PNC}}$$
$$= 2(T_{\text{NAV}}(B, \text{CTS}) - T_{\text{CO-PNC}}) - T_{\text{SIFS}} + T_{\text{ACK-PNC}}. \quad (9)$$

Eqs. (7)–(9) show that all the possible NAV lengths in the CO-PNC frame can be calculated based on the NAV lengths in the CTSs sent by the source nodes.

### 4.3.4 DATA

When only one node transmits data, the data frame does not update the NAV, and hence, the NAV length can be set to $T_{\text{NAV}}(A, \text{DATA}_{A\text{-only}}) = T_{\text{NAV}}(B, \text{DATA}_{B\text{-only}}) = 0$. When both nodes transmit data, the headers of the partly superposed frames are separately decoded, and the NAV timer is updated at the time the header has been completely received. Hence, nodes $A$ and $B$ respectively set the NAV length in its data frame to cover the remaining time used for data exchange after the header has been sent. Specifically, node $A$ sets the NAV length in its data frame to

$$T_{\text{NAV}}(A, \text{DATA})$$
$$= T_{\text{NAV}}(R, \text{CO-PNC}) - T_{\text{SIFS}} - T_{\text{PHY-Hd}} - T_{\text{MAC-Hd}}, \quad (10)$$

and node $B$ sets the NAV length in its data frame to

$$T_{\text{NAV}}(B, \text{DATA})$$
$$= T_{\text{NAV}}(R, \text{CO-PNC}) - 2T_{\text{SIFS}} - T_{\text{PHY-Hd}} - T_{\text{MAC-Hd}} - T_{\text{DATA}}(B). \quad (11)$$

## 4.4 Using Conventional Transmission Schemes

PNC-MAC operates in the same way as the IEEE 802.11 MAC protocol when conventional unicast packets are transmitted. When CNC is used, reliable broadcasting as proposed in [16] is employed to transmit the coded packets to the destinations, as well as the packets with overhearing opportunity to the relay and the opportunistic listeners.

## 5 QUEUING AND RELAYING METHOD SELECTION

First-in, first-out (FIFO) queuing is widely used in communication networks [27], which is a simple method of maintaining fairness among packets. At a sender with a FIFO queue, packets are sent on a first-come, first-served (FCFS) basis, i.e. the packets that arrive earlier at the queue are sent earlier. In PNC-MAC, we also intend to send packets on the FCFS basis. However, packet exchange with PNC is initiated by the relay, and not the source node. Therefore, we need to develop a specific queuing method for PNC-MAC, so that nodes can determine which packet should be sent. Meanwhile, the proposed PNC-MAC protocol concurrently supports PNC, CNC and conventional relaying schemes. Hence, we also need to develop a scheme to select the appropriate relaying method. In this section, we first discuss how to manage a specific queue used for the PNC-MAC protocol, and then focus on how to select the packet to send and its relaying method.

## 5.1 Managing the Queue

In PNC-MAC, each node manages two sender queues, which we respectively refer to as *actual queue* and *virtual queue*.

### 5.1.1 Elements in the Actual Queue

The actual queue stores the actual data packets that remain to be sent by the node, as in conventional communication networks. Each element in the actual queue contains the following fields.

1. Data packet, including its next and second hop addresses, which can be obtained from the routing information.
2. Time for which the packet has remained in the queue, i.e. from the time the packet enters the queue to present, which is denoted by $T_{\text{q}}(p)$, where $p$ represents the data packet.
3. Time for which the packet has remained in the queue of the packet's previous hop (if applicable), i.e. from the time the packet enters the queue of its previous hop to the time the packet exits the queue of its previous hop (and sent to the current node), which is denoted by $T_{\text{q-prev}}(p)$.

The latter two fields are used for selecting the packet to be transmitted and its relaying method, which will be discussed in Section 5.3.

### 5.1.2 Elements in the Virtual Queue

The virtual queue does not store actual data packets. It stores virtual data packets which contain some essential information of the packets in the actual queue of the node's neighbors. Issues on how this information is transferred will be explained in Section 5.1.4. The following fields are contained in every element in the virtual queue.

1. Virtual data packet, including: 1) previous hop, i.e. the node which currently has the corresponding actual packet in its actual queue; 2) next hop, i.e. the second hop of the corresponding actual packet; 3) length of the packet.
2. Time for which the corresponding actual packet has remained in the actual queue of the previous hop, denoted by $T_{\text{vq-prev}}(p_v)$, where $p_v$ represents the virtual packet, which corresponds to the actual packet $p$, and we have $T_{\text{vq-prev}}(p_v) = T_{\text{q}}(p)$.

The elements in the virtual queue are decreasingly ordered by $T_{\text{vq-prev}}$, i.e. the element with the largest $T_{\text{vq-prev}}$ is at the front of the queue.

### 5.1.3 Relationship between the Actual and Virtual Queues

To avoid unnecessary memory occupation, information of only a small subset of the packets in the actual queue of the node's neighbors is stored in the virtual queue. We have the following rules that connect the virtual queue of the node (e.g. *R*) with the actual queue of the node's neighbor (e.g. *A*).

*Rule 1:* All the virtual packets in the virtual queue of node *R* correspond to the packets in the actual queue of node *A* whose next hop is *R*, i.e. only information on packets that are expected to be sent to node *R* is stored in the virtual queue of node *R*.

*Rule 2:* At most one virtual packet with a specific previous hop (node *A*) and a specific next hop (e.g. node *B*) is stored in the virtual queue of node *R*.

*Rule 3:* The virtual packet (with previous hop *A* and next hop *B*) that is stored in the virtual queue of node *R* corresponds to the first packet in the actual queue of node *A* with next hop *R* and second hop *B*.

Note that the previous hop of the virtual packet corresponds to the node which currently has the actual packet in its actual queue, and the next hop of the virtual packet corresponds to the second hop of the actual packet. Rules 1–3 ensure that no redundant information is stored. An example of the relationship between the actual and virtual queues is given in Fig. 4.

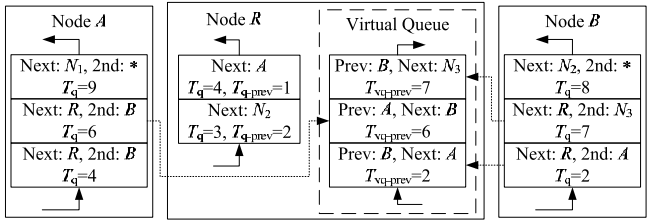

Fig. 4. Example of the relationship between the actual and virtual queues. The variables $N_1$, $N_2$ and $N_3$ denote arbitrary nodes which are different from *A*, *B* and *R*, a star (*) represents any node.

### 5.1.4 Transferring Queue Information to Neighbors

Information on packets in the node's actual queue is transferred to the node's neighbors in the data and ACK frames, so that the neighboring nodes can update their virtual queue.

The data frame carries information of the next packet, which has the same next hop and second hop as the data frame itself, in the actual queue. The same set of information as in each element of the virtual queue is contained as additional control information in the data frame. If the data frame has no second hop, the fields carrying information of the next packet are invalid. If there is no other packet in the queue with the same next and second hops, the length of the next packet is set to zero. As discussed in Section 5.1.1, the actual queue stores the time the packet has remained in the queue of its previous hop, which will be used in our later discussions.

Hence, the data frame also contains the time for which the packet itself has remained in the queue. Detailed frame formats will be discussed in Section 6.

After successfully receiving a data packet, the receiver sends an ACK frame carrying information of the next packet in the actual queue with the same next and second hops as the received data packet (the next and second hops are already updated by the receiver). The purpose of this setting is to update the virtual queue of the *next* hop (and hence, not the node that is being acknowledged) of the received packet. After updating the virtual queue, the next hop may initiate a PNC transmission if it finds it appropriate.

## 5.2 Checking for and Maintaining PNC Opportunity

By introducing the virtual queue, each node can judge whether it can initiate a PNC transmission to exchange packets for two of its neighboring nodes, according to the packets in its virtual queue. More explicitly, if there exist packets $p_v$ and $p_v'$ in the virtual queue satisfying: the previous hop of $p_v'$ is equal to the next hop of $p_v$, and the next hop of $p_v'$ is equal to the previous hop of $p_v$, then the actual packets corresponding to $p_v$ and $p_v'$ can be exchanged with PNC. The packet $p_v'$ is called the *reverse packet* of $p_v$ (and conversely) in our further discussions. For instance, in Fig. 4, the second and the third packets in the virtual queue of node *R* are reverse packets of each other.

Because PNC is initiated by the relay, the source nodes may still contend with the relay to access the channel and send their packets to the relay, even though there is opportunity to perform PNC. This channel contention is unnecessary and reduces the PNC opportunity. Therefore, if a node (e.g. *R*) senses that there is opportunity to perform PNC for its neighboring nodes (e.g. *A* and *B*), it notifies nodes *A* and *B* of PNC opportunity with a one-bit flag set in the data frames (whose previous hops are *B* and *A*, respectively) it sends to *A* and *B*. After node *A*, for instance, receives the data frame, it sets a flag to indicate that packets with the corresponding next hop (*R*) and second hop (*B*) should wait for PNC request from node *R*. These packets will not be sent, unless they are requested for a PNC transmission. The wait-for-PNC flag in node *A* will be cleared if at least one of the following conditions is satisfied.

1. No PNC request has been received during a timeout period $T_{\text{PNC-wait}}$.
2. When performing PNC, node *B* indicates in its header that it has no more packets (with the same next and second hops) in the actual queue to send.
3. Node *R* notifies node *A* that the wait-for-PNC flag should be cleared. This case can happen if node *R* receives at least one CTS indicating that the corresponding source node has no packet to send or if node *R* senses that there is no PNC opportunity.
4. Node *A* finds no other packet with the corresponding next hop and second hop addresses in its actual queue.

## 5.3 Selecting the Packet to Transmit and Its Relaying Method

According to the throughput gains between different relaying methods, the priority of different relaying schemes can be classified in the order as: PNC (highest), CNC, conventional relaying (lowest). A node first checks for PNC opportunity. If there is no opportunity to perform PNC, it switches to other relaying schemes according to the priority order. Meanwhile, in PNC-MAC, packet transmission is scheduled so that the FCFS basis is satisfied to the best effort, which means that queuing issues need to be considered. To consider both the actual and virtual queues, we propose the following rules.

*Rule 4:* When using CNC or conventional relaying, the first packet in the actual queue (denoted by $p$), excluding the packets that are waiting for PNC request, must be sent. When CNC can be performed, $p$ is coded with other packets in the queue before being sent.

*Rule 5:* PNC will be initiated if there exists a packet $p_v$ in the virtual queue and the following conditions are satisfied: 1) $T_{vq\text{-}prev}(p_v) \geq T_q(p)+T_{q\text{-}prev}(p)$; 2) there exists a reverse packet $p_v'$ of $p_v$ in the virtual queue. If there is more than one packet in the virtual queue that satisfies condition 1), the packet at the front of the virtual queue is considered first. By this means, we achieve fairness among nodes.

**Algorithm 1** Selecting the Packet to Transmit and Its Relaying Method
---
1: UsePNC = false
2: $p$ = Front_packet_of_actual_queue
3: **while** ($p$ != null **and** wait_for_PNC($p$))
4:     $p = p \rightarrow$ next
5: **end while**
6: $p_v$ = Front_packet_of_virtual_queue
7: **while** ($p_v$ != null **and** ($T_{vq\text{-}prev}(p_v) \geq T_q(p)+T_{q\text{-}prev}(p)$ **or** $p$ == null))
8:     **if** reverse packet $p_v'$ of $p_v$ exists in virtual queue
9:         UsePNC = true
10:         **break**
11:     **end if**
12:     $p_v = p_v \rightarrow$ next
13: **end while**
14: **if** UsePNC==true
15:     Schedule to request PNC to send the actual packets corresponding to $p_v$ and $p_v'$
16: **elseif** ($p$ != null)
17:     **if** packets that can be encoded with $p$ exist in actual queue
18:         Encode $p$ with the corresponding packets
19:         Schedule to send the coded packet
20:     **else**
21:         Schedule to send $p$
22:     **end if**
23: **end if**

Condition 1) in Rule 5 indicates that we link the actual queue and the virtual queue by considering the elapsed time starting from the time the packet enters the queue of its previous node. By this means, we have the same time measure for packets in the actual queue and the virtual queue, and the FCFS basis is achieved. Condition 2) in Rule 5 ensures that there is opportunity to perform PNC.

The algorithm for selecting the packet to transmit and the relaying method of the corresponding packet is shown in Algorithm 1. In the algorithm, PNC is first assumed to be unused, and the first packet $p$, which is not waiting for PNC request, in the actual queue is found. Then, Rule 5 is checked for packets in the virtual queue. If there is a packet in the virtual queue for which Rule 5 is satisfied, PNC is requested. Otherwise, the packet $p$ is sent. If there is coding opportunity, $p$ is sent with CNC. Otherwise, it is sent as a regular packet.

It is also worth to notice that the relaying method selection scheme is also a distributed solution to selecting appropriate relay nodes in a multi-hop wireless network which supports PNC. With the proposed scheme, each node judges itself whether it should act as a relay and perform PNC, according to the queue statuses.

## 5.4 The Operation Process

In this subsection, we discuss the operation process of PNC-MAC, with consideration of the queuing and relaying method selection issues.

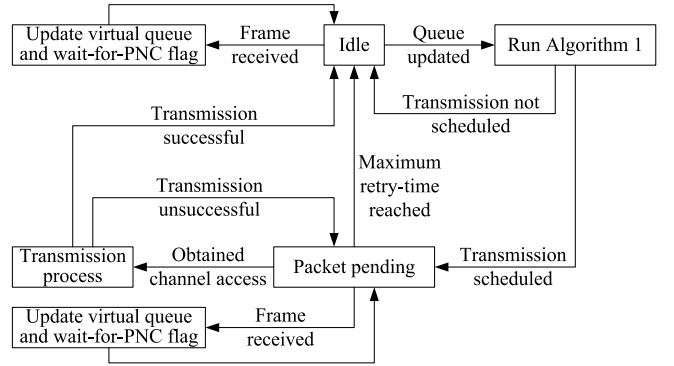

Fig. 5. State-transition diagram of PNC-MAC.

The state-transition diagram in Fig. 5 shows how PNC-MAC works. After a node has been initialized, it enters the idle state. When either the actual or the virtual queue is updated, Algorithm 1 is run, to select the packet to transmit and its relaying method. If transmission (including PNC, CNC or conventional transmission schemes) is scheduled in Algorithm 1, the node enters the packet pending state, to contend for channel access. Otherwise, the node returns to the idle state. In both of the idle and the packet pending states, when a frame is received, which carries information for updating the virtual queue or the wait-for-PNC flag, the corresponding updates are performed, and afterwards, the node returns to the state before receiving the frame. In the packet pending state, when the node obtains channel access, it starts the transmission process. When the transmission is successful, the node returns to the idle state and checks for new transmissions by running Algorithm 1 (note that if a transmission is successful, either the actual or the virtual queue will be updated). A PNC transmission is regarded as successful if at least one node has successfully received a packet. This corresponds to the case where only one source node transmits a packet and the relay has successfully received it, or where two source nodes transmit packets, which is the ordinary case, and at least one packet is received by the destination(s). When the transmission is unsuccessful, the node returns to the packet pending state, to retry to send the pending packet. When the maximum retry-time has been reached, the node flushes the pending packet from the queue (when using PNC, both pending virtual packets are flushed from the virtual queue) and returns to the idle state.

## 6 FRAMING

In this section, we discuss the framing of the proposed PNC-MAC protocol. The frames are designed to contain the necessary information as discussed in Sections 4 and 5. Fig. 6 summarizes the formats of several frames of PNC-MAC which are modified according to the IEEE 802.11 standard.

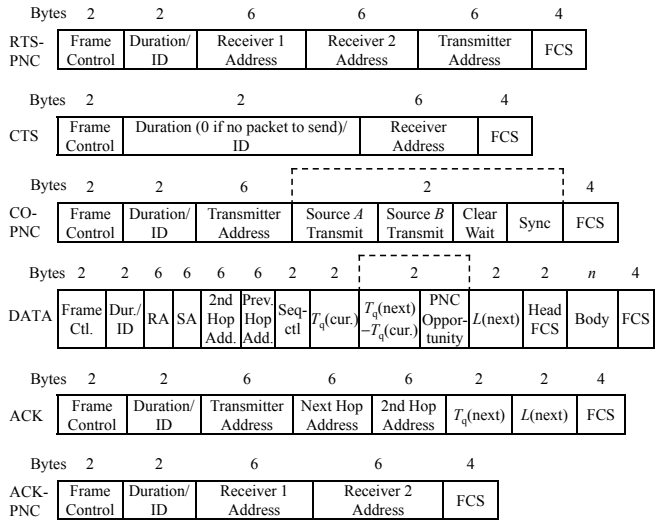

Fig. 6. Frame formats of the PNC-MAC protocol.

### 6.1 RTS-PNC

The RTS-PNC frame is similar with the RTS frame of the IEEE 802.11 MAC, except that addresses of two receivers are included.

### 6.2 CTS

The CTS frame has the same format as the IEEE 802.11 CTS frame. If a source node has no packet to send, it sets the duration field to zero. In this case, the NAV timers of other nodes are not updated, and the relay can also get to know that the source node has no packet to send by this means.

### 6.3 CO-PNC

The CO-PNC frame contains the transmitter address, i.e. the address of the relay node, which is an identifier of the current PNC session. After the transmitter address, a two-bytes' control information is followed, which include whether the source nodes $A$ and $B$ have data to transmit and whether the wait-for-PNC flag in the source nodes should be cleared. The remaining bits can be used for synchronization, as discussed in Section 4.1.

### 6.4 DATA

Except for the receiver address (RA) and the source address (SA) in the IEEE 802.11 data frame, the data frame of PNC-MAC also contains the second hop and the previous hop addresses. Meanwhile, for updating the queue information as discussed in Section 5, it also contains the time $T_q$(cur.) for which the packet has remained in the actual queue of the source node, the time $T_q$(next) for which the next packet (with the same next and second hops) has remained in the queue, and the length $L$(next) of the next packet. To make the frame more compact, rather than transmitting $T_q$(next), the time offset between $T_q$(next) and $T_q$(cur.) is transmitted, which occupies 15 bits out of two bytes. The remaining bit is used for setting the wait-for-PNC flag of the receiver, as discussed in Section 5.2.

### 6.5 ACK

In the ACK frame, we use the transmitter address (rather than the receiver address as in the IEEE 802.11 standard) to indicate which data-exchange session is to be acknowledged. The transmitter address corresponds to a unique receiver address, because a node can only send CTS to one specific source node. Using the transmitter address is for considerations of updating the virtual queue. Also for this consideration, the ACK frame contains the next hop and second hop addresses of the received packet, as well as $T_q$(next) and $L$(next).

### 6.6 ACK-PNC

The ACK-PNC frame contains two receiver addresses that are to be acknowledged.

### 6.7 Broadcast Packets for CNC

When using CNC, the RTS and the data frames contain addresses of all destinations, to achieve reliable broadcasting.

## 7 PERFORMANCE EVALUATION

In this section, we evaluate the performance of the proposed PNC-MAC protocol through simulations. The simulation program is jointly built on MATLAB and C. It is discrete event driven and has detailed physical-layer modeling. The simulation program first simulates the signal transmission process. According to the received signal, the bit-error-rates (BERs) and packet-error-rates at the receiver are calculated. Then, the simulator decides whether a packet is successfully received, based on the packet-error-rate. The physical-layer model is based on digital communication theories [30] and its details are described in Appendix A.

### 7.1 Simulation Setup

In our simulations, we use the IEEE 802.11 direct-sequence spread spectrum (DSSS) physical-layer with 1 Mbps data rate. The transmission power is set to 3 dBm, the background noise density is –174 dBm/Hz, and the noise figure is 6 dB. With these settings, the interference-free communication range of nodes is approximately 250 m. The network traffic is modeled with User Datagram Protocol (UDP) flows with packet size of 1000 bytes. The sizes of the actual and virtual queues at the sender are both 50 packets. Unless specified, the timeout of the wait-for-PNC flag $T_{PNC\text{-}wait}$ is set to 1 s, and the receiver's clear channel assessment (CCA) sensitivity is set to –100 dBm.

The CCA sensitivity is set lower than the interference-free received signal strength (RSS) requirement of successful transmission (that is approximately –93.2 dBm for a packet loss rate of 1%, which can be evaluated according to Appendix A). In other words, the sensing range is set to a larger value than the transmission range. Because the interference range in wireless transmissions is generally larger than the transmission range [28], this setting ensures that nodes can sense the signals that may interfere with the transmission (although they may not be able to receive an error-free packet that the signal carries) and avoid accessing the channel when such an interference signal is present. Meanwhile, the CCA sensitivity of –100 dBm is achievable in practical devices [29], and we will show in Section 7.3.2 that this setting brings the highest throughput for PNC-MAC.

We compare the performance of PNC-MAC with MAC protocols that do not support PNC. These protocols include a MAC protocol with CNC support (referred to as CNC-MAC), as discussed in [16], and the IEEE 802.11 MAC protocol [11]. The performance is first evaluated in simple network topologies (as shown in Fig. 7) in Sections 7.2 and 7.3, and then evaluated in random topologies in Section 7.4. The simulation time is 50 s. Each simulation was run with 10 different random seeds to evaluate the overall performance.

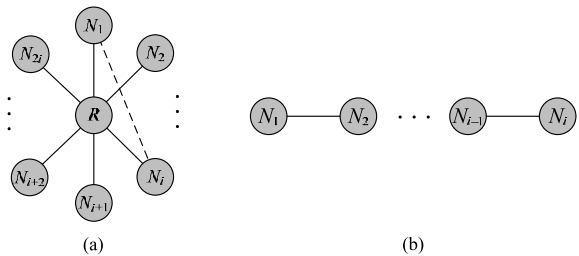

Fig. 7. Simple topologies: (a) wheel topology, (b) line topology.

## 7.2 Wheel Topology

The wheel topology is shown in Fig. 7(a), in which nodes exchange packets through a common relay $R$. Opposite nodes exchange packets with each other, i.e. the data flows in Fig. 7(a) are $N_1 \leftrightarrow N_{i+1}$, $N_2 \leftrightarrow N_{i+2}$, etc. The nodes $N_1, N_2, \ldots, N_i, \ldots, N_{2i}$ are uniformly placed in a circle centered at $R$. The radius of the circle is set so that the non-opposite nodes (e.g. nodes $N_1$ and $N_i$ in Fig. 7(a)) can communicate with each other and the opposite nodes cannot communicate with each other. The maximum value of the radius is 150 m. The nodes, except for the relay, are backlogged, i.e. they always have packets to send. After a packet has been sent to the relay, a new packet is generated and put into the queue. The queue of each source node always keeps two packets that are waiting to be sent. Fig. 8 shows the resulting average throughput per end node and Fig. 9 shows the end-to-end delay, with different number of nodes in the circle.

It can be observed from Fig. 8 that PNC-MAC has the highest throughput. When there are two nodes in the circle, the throughput gain of PNC-MAC over CNC-MAC is 1.48, which approaches the theoretical result 1.5, as discussed in Section 1 and [17]. When there are 10 nodes in the circle, the corresponding throughput gain is 4.75. The throughput gain increases with the number of nodes. The reason is that when using PNC-MAC in the wheel topology, the relay acts as a coordinator which coordinates transmissions of the source nodes, and nodes do not perform channel contention. When using CNC-MAC or 802.11, the channel contention is becomes intense with the increasing node number, which results in frequent collisions and reduces the throughput.

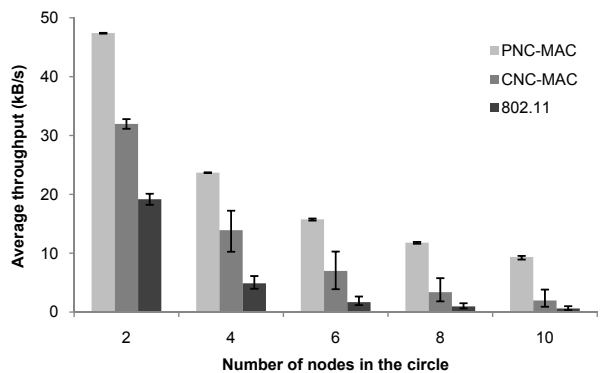

Fig. 8. Average throughput vs. number of nodes in the wheel topology. The error bars indicate the maximum and minimum values.

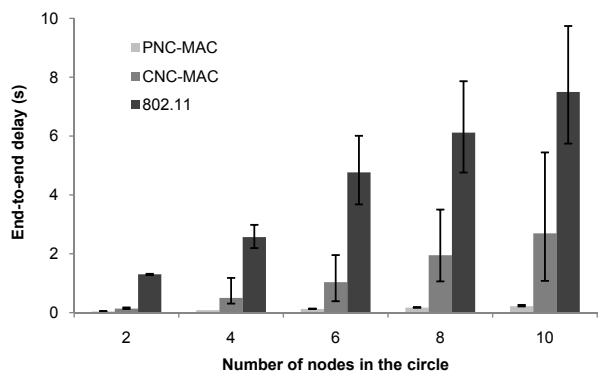

Fig. 9. End-to-end delay vs. number of nodes in the wheel topology. The error bars indicate the maximum and minimum values.

The end-to-end delay of PNC-MAC in the wheel topology is the lowest compared with the other two protocols, as shown in Fig. 9. The reason is that when using PNC, the data packets do not enter the queue of the relay. Instead, they are forwarded to the destinations immediately after they have been received by the relay. By this means, additional delaying, which would occur in the actual queue of the relay, is reduced.

The variations of the throughput and the end-to-end delay among different simulation instances and different nodes are lowest when using PNC-MAC, because in this case, the transmissions are coordinated by the relay and the randomness due to channel contention is reduced. Fairness is achieved by the introduction of virtual queues, which makes sure that the packet which has remained in the queue for the longest time will be sent.

## 7.3 Line Topology

As shown in Fig. 7(b), in the line topology, nodes are placed in a line. The distance between neighboring nodes in the line is 150 m. The end nodes, i.e. $N_1$ and $N_i$ in Fig. 7(b), exchange packets and are backlogged.

### 7.3.1 Number of Nodes

The total throughput and average end-to-end delay at different number of nodes are shown in Figs. 10 and 11. It can be observed from Fig. 10 that PNC-MAC outperforms the other two protocols, with an average throughput gain of 1.48 over CNC-MAC. The end-to-end delay of PNC-MAC is similar with the delay of CNC-MAC, as shown in Fig. 11. PNC-MAC does not bring much delay improvement because the increased throughput allows more packets to enter the network, and consequently, packets may have to wait for a longer time in the queue. However, the delays of PNC-MAC and CNC-MAC are both substantially lower than the delay of 802.11.

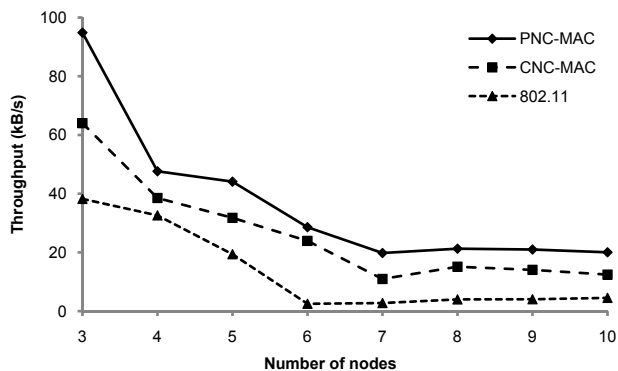

Fig. 10. Total throughput vs. number of nodes in the line topology.

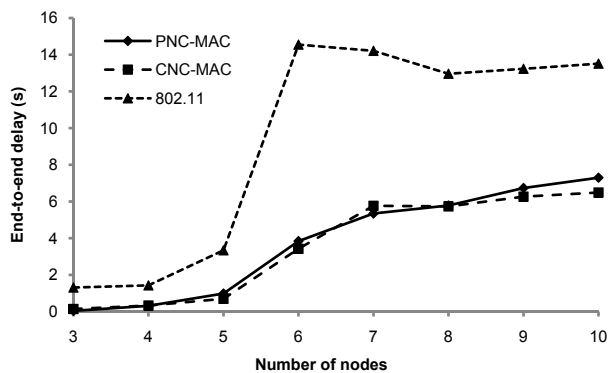

Fig. 11. End-to-end delay vs. number of nodes in the line topology.

### 7.3.2 CCA Sensitivity

The impact of the CCA sensitivity on the network performance is studied in this subsection, where the number of nodes is set to 10.

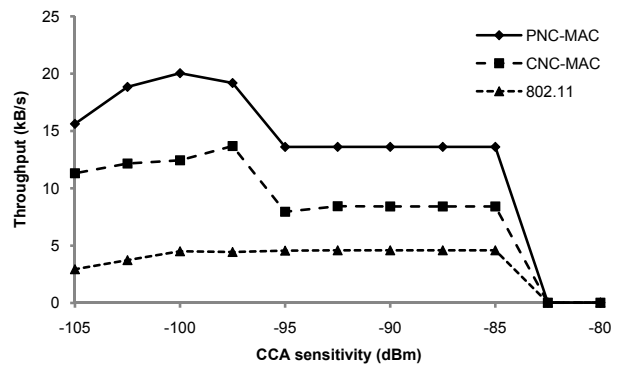

Fig. 12. Total throughput vs. CCA sensitivity in the line topology.

It can be observed from Fig. 12 that, when the CCA sensitivity is greater than or equal to –82.5 dBm, the throughputs of all the protocols are zero. This corresponds to the case where nodes cannot sense the transmission of any other node[1] and collision occurs very frequently. From –95 dBm to –85 dBm, the throughputs remain approximately unchanged, which corresponds to the case where every node can only sense its one-hop neighbors.

PNC-MAC has the highest throughput when the CCA sensitivity is –100 dBm, and CNC-MAC achieves the highest throughput at –97.5 dBm. At both values each node can sense the transmission of nodes that are two hops away. The difference between the optimal CCA sensitivities is due to the bit-error-rate (BER) increment when performing PNC, as discussed in Appendix A.2, causing PNC-MAC less tolerable to interference than CNC-MAC.

The throughput of 802.11 is remains similar when the CCA sensitivity is between –100 dBm and –85 dBm, i.e. sensing the nodes that are two hops away does not bring much benefit for 802.11. The reason is that, with our simulation setup, only unicast transmissions are performed with 802.11. When nodes are able to sense those nodes that are two hops away, the network exhibits the exposed terminal problem, preventing nodes that would actually not cause a collision to transmit [11]. On the other hand, when nodes are only able to sense their one-hop neighbors, we have the hidden terminal problem, leading to destructive collisions. For unicast transmissions, the exposed and hidden terminal problems take similar effects to the network, and we observe a similar throughput for the two cases.

However, sensing two hops is beneficial for PNC-MAC and CNC-MAC, which utilize multicast

---

[1] According to the simulation setup and the channel model, the (interference-free) RSS from a neighboring node is $10\log_{10}(10^{(3/10)}/150^4)$ = –84.0 dBm, the RSS from a node that is two hops away is $10\log_{10}(10^{(3/10)}/300^4)$ = –96.1 dBm, and the RSS from a node that is three hops away is $10\log_{10}(10^{(3/10)}/450^4)$ = –103.1 dBm.
11

transmissions and are more likely to exhibit collisions when hidden terminals are present.

When the CCA sensitivity is set lower, so that nodes that are three or more hops away can be sensed, the throughputs of all the protocols decrease, due to inefficient channel utilization.

Meanwhile, the throughputs vary more gradually when the CCA sensitivity is smaller than or equal to –97.5 dBm (compared with larger sensitivity values), because in this case, the CCA sensitivity is no longer strictly related to the number of hops that a node can sense. The addition of interference signals from various far-away nodes may also cause the RSS to be higher than the CCA sensitivity.

In terms of end-to-end delay, Fig. 13 shows that, as the CCA sensitivity decreases, the delay of 802.11 increases. This is due to the low channel utilization when the CCA sensitivity is low. The delays of PNC-MAC and CNC-MAC do not vary significantly with the CCA sensitivity, because the decreased channel efficiency is balanced by PNC or CNC opportunities.

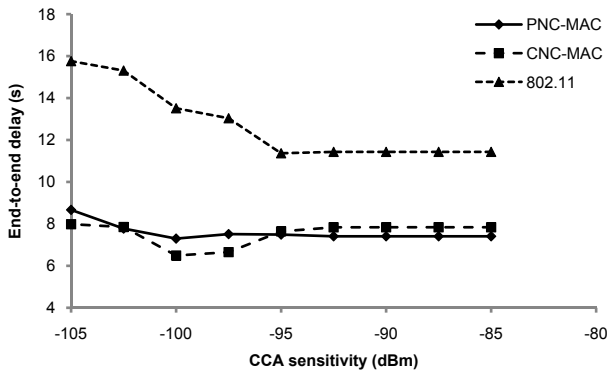

Fig. 13. End-to-end delay vs. CCA sensitivity in the line topology.

### 7.4 Random Topology

The random topology under consideration contains 40 nodes randomly distributed in a 1000 × 1000 m² network area. Among these 40 nodes, 20 nodes are randomly selected to exchange packets in node-pairs, i.e. 10 bi-directional flows are configured. The packets are generated according to a Poisson arrival process.

#### 7.4.1 Packet Rate

We first consider the impact of the packet rate on the network performance. The throughput and end-to-end delay for different packet rates are shown in Figs. 14 and 15.

It can be observed from Fig. 14 that PNC-MAC outperforms CNC-MAC with a maximum throughput gain of 1.52 and an average throughput gain of 1.33. The highest throughputs of CNC-MAC and 802.11 are achieved when the packet rate is 5 packets/s. Their throughputs decrease as the packet rate increases beyond that value. This is a common observation for multi-hop networks with UDP flows [4], [22], due to the intense channel contention at high packet rates and the lack of load balancing scheme.

The throughput of PNC-MAC also reaches its local maxima when the packet rate is 5 packets/s. At higher packet rates, the throughput remains similar, even with a slight increase. This is because the inherent operation mode of PNC-MAC can also balance the load of the network, by setting the wait-for-PNC flag for nodes with PNC opportunity.

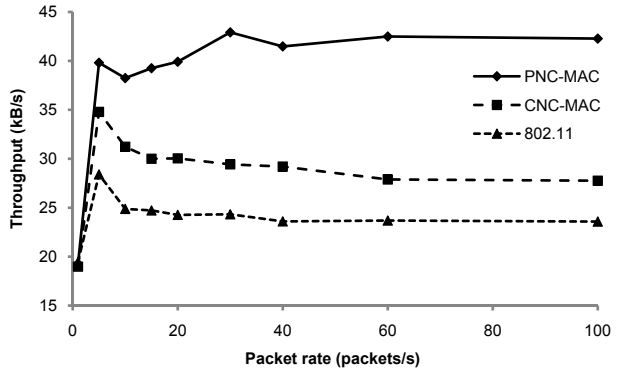

Fig. 14. Total throughput vs. packet rate in the random topology.

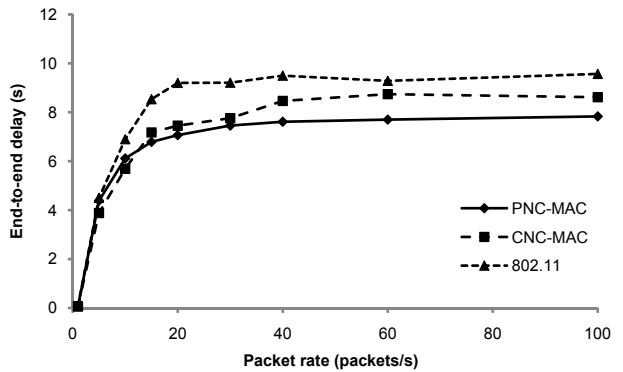

Fig. 15. End-to-end delay vs. packet rate in the random topology.

The delay of PNC-MAC is similar with the delay of CNC-MAC when the packet rate is low, as shown in Fig. 15. For higher packet rates, the delay of PNC-MAC is lower than the delay of CNC-MAC. The reason is that, when the packet rate is high, nodes have to wait for longer time in the queue before being transmitted. PNC becomes beneficial in this case because, at the relay, the superposed packets are not put into the queue and are immediately forwarded to the destinations. The wait-for-PNC flag also reduces the number of the contending nodes to some extent, which can decrease the delay.

#### 7.4.2 Wait-for-PNC Timeout

In this subsection, we study the impact of the timeout $T_{\text{PNC-wait}}$ of the wait-for-PNC flag on the network performance. The packet rate is set to 5 packets/s, where CNC-MAC and 802.11 have their highest throughputs.

Fig. 16 shows that the throughput increases with the wait-for-PNC timeout. This is because the wait-for-PNC flag is updated (i.e. set or cleared according to the queue statuses) in real-time, as described in Section 5; and messages for updating the wait-for-PNC flag is carried in the corresponding frames, as described in Section 6. Therefore, for a static network, a long wait-for-PNC timeout is beneficial to the performance, unless there exists a long-period packet loss, which is not likely to happen because we set the CCA sensitivity to a value that can capture the most interference. We consider static networks because routing issues need to be considered in a network with mobility, which is beyond the scope of this paper. However, in a network containing mobile nodes, the wait-for-PNC flag can be cleared after each time a link failure is detected. Link failure detection (and routing table updating) schemes are integrated in many routing protocols [27].

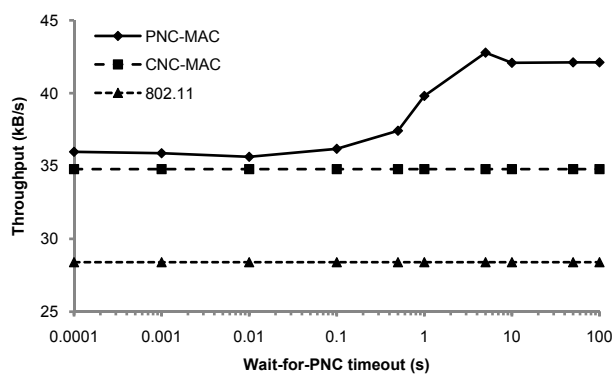

Fig. 16. Total throughput vs. wait-for-PNC timeout in the random topology.

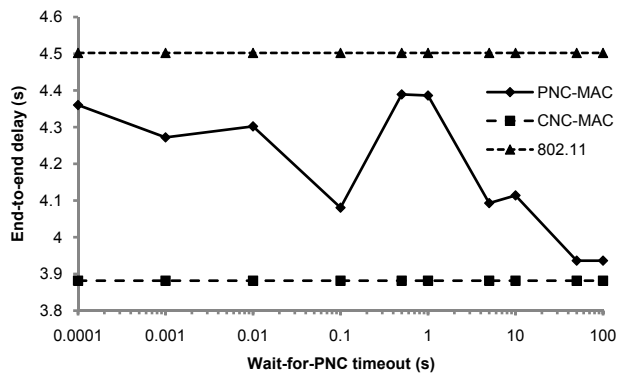

Fig. 17. End-to-end delay vs. wait-for-PNC timeout in the random topology.

The end-to-end delay for different wait-for-PNC timeout values is shown in Fig. 17. We can observe that the differences between the delays of the three protocols are within 1 s. The delay of PNC-MAC reaches its highest value when the timeout is around 1 s. One possible reason for this fact is, when the timeout is around 1 s, some packets first wait for the PNC request and then start to contend the channel, because the timeout event has occurred and the relay has not yet requested PNC transmission. This causes additional delay as well as intense channel contention between the source nodes and the relay. The performance is improved when the timeout value is larger, in which case nodes wait for the PNC request for a sufficient time and do not contend the channel in the meantime.

## 8 CONCLUSIONS

In this paper, we have proposed the PNC-MAC protocol, which extends the IEEE 802.11 MAC protocol to support PNC, as well as CNC and conventional relaying methods. With PNC-MAC, the node decides whether PNC, CNC, or conventional transmission should be initiated, according to its actual and virtual queues. A PNC transmission is initiated by the relay, which also acts as a coordinator that coordinates the whole packet exchange process. The simulation results show that the proposed PNC-MAC protocol brings throughput improvement in various scenarios compared with the protocols that do not support PNC, while maintaining a similar delay as CNC-MAC. It follows that PNC-MAC is beneficial for throughput-sensitive applications of wireless networks.

We have considered a network with bi-directional flows in this paper. Scenarios with unidirectional flows and the support of opportunistic listening for PNC are also worth investigating in the future. In order to support opportunistic listening, the protocol may need to estimate the channel status and predict future channel conditions, because the source and destination nodes do not overlap. It is also interesting to investigate the possibility of combining PNC-MAC with MAC protocols that support SIC, to improve the performance when unidirectional flows exist.

## ACKNOWLEDGMENT

This work was supported in part by the Fundamental Research Funds for the Central Universities under Grant Nos. N100404008, N110204001, and N110204003, the National Natural Science Foundation of China (Nos. 61172051, 61070162, 71071028, and 70931001), the Fok Ying Tung Education Foundation (No. 121065), and the Specialized Research Fund for the Doctoral Program of Higher Education (Nos. 20110042110023, 20110042110024, and 20120042120049).

**Shiqiang Wang** received the BEng and MEng degrees from Northeastern University, China, in 2009 and 2011, respectively. He is currently working toward the PhD degree in the Department of Electrical and Electronic Engineering, Imperial College London, United Kingdom. His research interests include network coding, protocol design, optimization, and prototyping for wireless networks. He has a dozen scholarly publications in international journals and conferences. He served on the program committee of IEEE VTC 2012-Fall, 2013-Spring, and 2013-Fall.

**Qingyang Song** received the PhD degree in telecommunications engineering from the University of Sydney, Australia. She is an associate professor at Northeastern University, China. She has authored more than 30 papers in major journals and international conferences. These papers have been cited more than 500 times in scientific literature. Her current research interests are in radio resource management, network coding, cognitive radio networks, and cooperative communications.

**Xingwei Wang** is a Professor of the School of Information Science and Engineering, Northeastern University. He received his BS, MS and PhD degrees from Northeastern University in 1989, 1992 and 1998 respectively. His research and teaching interests focus on computer networks, grid computing and information security. He has undertaken dozens of funded research projects supported by National Natural Science Foundation of China and Ministry of Science and Technology of China, etc. He has published over 200 scholarly publications in IEEE Transactions on Communications, IEEE/KICS Journal of Communications and Networks, IEEE Transactions on Vehicular Technology, ACM Operating Systems Review, Computer Networks, Computer Communications, etc. He has won two National Science and Technology Progress Awards.

**Abbas Jamalipour** (S'86-M'91-SM'00-F'07) holds a PhD from Nagoya University, Japan, and the Chair Professor of Ubiquitous Mobile Networking at the University of Sydney, Australia. He is a Fellow of IEICE and IEAust, an IEEE Distinguished Lecturer and a Technical Editor of several scholarly journals. He has been an organizer or chair of many international conferences including IEEE ICC and GLOBECOM and was the General Chair of the IEEE Wireless Communications and Networking Conference (IEEE WCNC 2010). He is the Chair of Communications Switching and Routing Technical Committee, Vice Director of Asia Pacific Board, and a voting member of Conference Boards, Education Board, and Online Contents of the IEEE Communications Society. He is the recipient of several prestigious awards such as the 2010 IEEE ComSoc Harold Sobol Award for Exemplary Service to Meetings & Conferences, the 2006 IEEE ComSoc Distinguished Contribution to Satellite Communications Award, and the 2006 IEEE ComSoc Best Tutorial Paper Award.

## Appendix A: Physical-Layer Model

### A.1 Signal Transmission

Let $S$ denote the set of nodes that are currently transmitting data, suppose node $j$ intends to receive the data sent by the nodes in the set $S'$, the received signal at node $j$ can be evaluated by

$$y_j = \sum_{i \in S'} h_{ij} x_i + \sum_{k \notin S', k \in S} h_{kj} x_k + z_n, \quad \text{(A-1)}$$

where $x_i$ denotes the signal transmitted by node $i$, $h_{ij}$ denotes the channel gain from node $i$ to node $j$, and $z_n$ is additive white Gaussian noise (AWGN) with power spectral density $N_0$.

In our simulations, the power gain of the channel between two nodes is evaluated by $|h_{ij}|^2 = 1/D_{ij}^4$, where $D$ is the distance between nodes $i$ and $j$ in meters. The phase of the channel gain $h_{ij}$ is uniformly distributed within $[0, 2\pi)$. The size of $S'$ is one for conventional transmissions; and it is two for PNC.

According to the Central Limit Theorem, the sum of independent random variables has approximately a Gaussian distribution. Hence, we assume that

$$\sum_{k \neq i, k \in S} h_{kj} x_k \sim CN(0, IT_s), \quad \text{(A-2)}$$

where $I$ is the total *destructive* interference power at the receiver node $j$, $T_s$ is the time-length of each symbol, and $CN(\mu, \sigma^2)$ denotes the circularly-symmetric Gaussian distribution with mean $\mu$ and variance $\sigma^2$ [30]. It follows that

$$\sum_{k \neq i, k \in S} h_{kj} x_k + z_n \sim CN(0, IT_s + N_0). \quad \text{(A-3)}$$

### A.2 Bit-Error-Rate

We use the same parameters as the IEEE 802.11 direct-sequence spread spectrum (DSSS) physical-layer with 1 Mbps data rate and differential binary phase-shift keying (DBPSK) modulation [11] in the simulations. The receiver operates under the minimum distance decision rule [30].

From digital communication theories [30] and (A-1), (A-2), and (A-3), the BER for DBPSK modulation can be evaluated by

$$P_{e\text{-DBPSK}} = 2Q\left(\sqrt{\frac{2E_s}{N_0 + IT_s}}\right), \quad \text{(A-4)}$$

where $E_s$ is the received energy per symbol. We have $E_s = P_T |h_{ij}|^2 T_s$, where $P_T$ is the transmission power of nodes.

When performing PNC, we use the DNF method with symbol-level synchronization, but without phase-level synchronization. In this case, each constellation point in the constellation of the superposed signal has a maximum of two neighbors that correspond to different symbols [7]. Hence, the BER of the DNF process, which is denoted by $P_{e\text{-DNF}}$, is constrained by

$$P_{e\text{-DBPSK}} \leq P_{e\text{-DNF}} \leq 2P_{e\text{-DBPSK}}. \quad \text{(A-5)}$$

Note that the DNF method with symbol-level synchronization is only one possible PNC realization method for PNC-MAC. We choose this method because it simplifies our analysis. The symbol-level synchronization requirement can be relaxed when using asynchronous DNF schemes [9].

As a conservative approach, we take the upper bound of (A-5) as the BER value of the DNF process in the simulations. We also use the lower value of $E_s$ (note that the values of $E_s$ of the two independent signal components of the superposed signal are generally different) for calculations.

The IEEE 802.11 DSSS uses an 11-bit Barker word as the spreading sequence [11]. The receiver first decodes the spread-coded sequence, with the aforementioned BERs. Then, it correlates the received sequence with the known Barker word, to obtain the original bits. Hence, the original bit is erroneous when at least six bits in the received sequence is erroneous, and for the original bits, the BER can be evaluated by

$$P_{e\text{-orig}} = \sum_{m=0}^{5} \binom{11}{m} (1 - P_{e\text{-spread}})^m P_{e\text{-spread}}^{11-m}, \quad \text{(A-6)}$$

where $P_{e\text{-spread}}$ is the BER of the spread-coded sequence, which can be equal to $P_{e\text{-DBPSK}}$ or $P_{e\text{-DNF}}$.

### A.3 Packet-Error-Rate

Without forward error correction schemes, a packet is erroneous when at least one bit in the packet is erroneous. Hence, the packet-error-rate can be evaluated by

$$P_{e\text{-pk}} = 1 - \prod_{l=1}^{L} [1 - P_{e\text{-orig}}(l)], \quad \text{(A-7)}$$

where $L$ denotes the packet length and $P_{e\text{-orig}}(l)$ denotes the BER of the $l$th original bit of the packet.

The simulation program drops packets at the same rate as the packet-error-rate.